\documentclass[pdflatex,sn-mathphys-num]{sn-jnl}

\usepackage{graphicx}%
\usepackage{multirow}%
\usepackage{amsmath,amssymb,amsfonts}%
\usepackage{amsthm}%
\usepackage{mathrsfs}%
\usepackage[title]{appendix}%
\usepackage{xcolor}%
\usepackage{textcomp}%
\usepackage{manyfoot}%
\usepackage{booktabs}%
\usepackage{ulem} 
\usepackage{algorithm}%
\usepackage{algorithmicx}%
\usepackage{algpseudocode}%
\usepackage{listings}%

\usepackage{bm}
\usepackage{lineno}
\def\ii{{\mathrm{i}}}

\usepackage{here}

\usepackage{booktabs}
\usepackage{diagbox}
\usepackage{textcomp}
\usepackage[table]{xcolor}

\theoremstyle{thmstyleone}%
\theoremstyle{thmstyletwo}%

\theoremstyle{thmstylethree}%

\begin{document}

\title[Article Title]{Single-pixel edge enhancement of object via convolutional filtering with localized vortex phase}


\author*[1]{\fnm{Jigme} \sur{Zangpo}}\email{jigmezangpo11@gmail.com}
\author[2]{\fnm{Hirokazu} \sur{Kobayashi}}
\author[2]{\fnm{Takumi} \sur{Jinushi}}
\author[1,3]{\fnm{Ryo} \sur{Yasuhara}}

\affil[1]{\orgdiv{National Institute for Fusion Science}, \orgaddress{\street{ 322-6 Oroshi-cho}, \city{Toki}, \postcode{5095292}, \state{Gifu}, \country{Japan}}}

\affil[2]{\orgdiv{Graduate School of Engineering}, \orgname{Kochi University of Technology}, \orgaddress{\street{185 Miyanokuchi, Tosayamada}, \city{Kami}, \postcode{610101}, \state{Kochi}, \country{Japan}}}

\affil[3]{\orgdiv{
The Graduate University for Advanced Studies}, \orgaddress{\street{SOKENDAI, 322-6 Oroshi-cho}, \city{Kami}, \postcode{509-5292}, \state{Toki}, \country{Japan}}}


\abstract{In this paper, we propose and numerically demonstrate an all-directional edge-enhanced microscopy based on single-pixel imaging using convolutional filtering with a localized vortex phase. The edge-enhanced microscope with a vortex filter is of particular interest for optical information processing, as it can highlight both amplitude and the phase edges of complex objects in all directions. While Fourier-domain single-pixel imaging has been proposed as a cost-effective approach for edge enhancement in non-visible wavelengths, it requires three or four times more measurements due to the need for three- or four-phase shifting to detect optical complex amplitudes in the Fourier domain. Our method implements convolutional filtering with a localized vortex phase in the spatial domain, eliminating the need for these extra measurements required by phase-shifting methods. Simulations demonstrate a high correlation coefficient of 0.95 between ideal and enhanced edges, offering significant improvements for edge enhancement in various invisible-wavelength imaging applications.}




\maketitle

\section{Introduction}
Microscopy is an essential tool in imaging research, and edge-enhanced microscopy with vortex filters has attracted particular attention for enhancing phase and amplitude object edges due to its capability for all-directional edge enhancement \cite{spiralPF2,spiralPF4,spiralPF7,spiralPF8,zangpo2023edge,zangpo2024isolation,zangpoedge}, whereas differential interference-contrast microscopy enhances the edges in only one direction\cite{lang1982nomarski}. The application of this technique is not limited to the visible range, but the edge enhancement of an object in the invisible wavelength is also crucial for near-infrared fluorescence and electronic circuit inspection through silicon semiconductors. One disadvantage of near-infrared imaging is that digital cameras, such as silicon-based charge-coupled devices (CCD) and complementary metal-oxide-semiconductors (CMOS), tend to be more expensive than cameras for the visible spectrum\cite{duarte2008single,lowcost}.

To address this issue, researchers have opted for a more cost-effective approach by utilizing single-pixel detectors with the reconstruction of two-dimensional images by correlation calculation\cite{duarte2008single, gibson2020single,edgar2019principles,ghostimaging}. So-called single-pixel imaging can detect a weak signal and image with a high speed\cite{sensitivity1, sensitivity2,speed1,speed2,highspeed}. In Refs. \cite{nonvisiblelight,phase4singlepixel1,phase4singlepixel2,phase4singlepixel3}, researchers used single-pixel imaging to capture complex amplitude distribution of objects using a four-step phase-shifting technique. Subsequently, researchers attempted to integrate single-pixel imaging into an edge-enhanced microscope, where Fourier single-pixel imaging was adopted to capture the edges of the phase object using three- or four-step phase shifting\cite{VPM4phase1, VPM4phase2,VPM4phase3,VPM3phase2}. The advantage of their approach is the absence of a vortex filter; instead, they utilized a digital micromirror device (DMD) to generate complex wavefronts through various modulation techniques. Although various methods such as error diffusion \cite{referee_1_errorDiffussion} (offering efficient pattern generation through quantization error distribution) and Direct Binary Search \cite{referee_1_DBS} (providing high-precision wavefront control) exist, these approaches face trade-offs between speed and fidelity. Most implementations employ a reference wave and an optical vortex with the assistance of a super-pixel technique\cite{superpixel1, superpixel2, superpixel3}, which is highly robust and easy to use while offering full spatial control over the phase and amplitude of a light field\cite{superpixel1,ampPhasewavefront}. While this method successfully detects and enhances the edges of the phase object, the drawback is that the three- or four-step phase shifting, requires three or four times as many single-pixel measurements for reconstruction. In contrast, the technique used for edge detection in Ref. \cite{computation_2} is a speckle-shifting ghost imaging method, where the gradient operation is applied directly to the illumination patterns rather than the object image, combined with morphology-based denoising algorithms to enhance edge clarity and reduce noise. However, this approach still requires multiple measurements and additional computational steps for noise reduction.

To reduce the number of single-pixel measurements, we propose a method, namely, a single-pixel edge-enhanced microscope via convolutional filtering with a localized vortex phase, which eliminates the extra measurements required by the phase-shifting method. Numerical simulations validate this approach, achieving a correlation coefficient of 0.95 with ideal edge enhancement, demonstrating its precision and effectiveness. Compared to speckle-shifting ghost imaging\cite{computation_1}, our method is more efficient, requiring fewer measurements, and computationally simpler, as it directly enhances the edges without the need for additional denoising steps, while maintaining high accuracy in edge detection.

In Section 2 we describe the proposed method, the single-pixel edge enhancement of an object, and the principle of the super-pixel method. In Sections 3, we present the numerical simulation results. Section 4 concludes the study.

\section{Single-pixel Edge Enhancement via Convolutional Filtering with Localized Vortex Phase}
\subsection{Principle of single-pixel edge enhancement via convolutional filtering}
\label{proposed method}
\begin{figure}[htp!]
 \centering
\includegraphics[width=\linewidth]{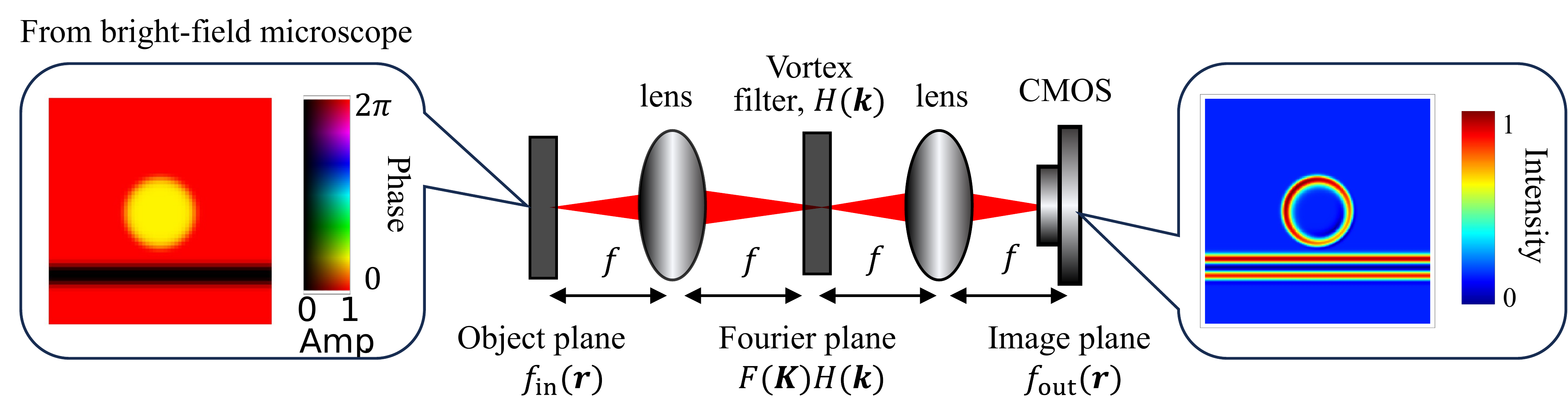}
\caption{The $4f$ system with a vortex filter for all-directional edge enhancement.}
\label{4fsystem}
\end{figure}

Figure \ref{4fsystem} shows the $4f$ system with a vortex filter to implement all-directional edge enhancement placed after the bright-field microscope. The $4f$ system consists of an object plane, Fourier plane and image plane. At the object plane, the object magnified by the bright-field microscope is of the phase-amplitude object (PAO) type, whose brightness (from 0 to 1) denotes amplitude and whose color denotes phase. In the object, a yellow circle and a horizontal black strip indicate the phase object and the amplitude object, respectively. The input object $f_\text{in}(\bm{r})$ magnified by the bright-field microscope undergoes the Fourier transformation by the first lens and the Fourier spectrum is given by  $F(\bm{k})=\mathcal{F}\{f(\bm{r})\}$, where $\mathcal{F}$ represents the Fourier transform, $\bm{r}=(x,y)$ represents the two-dimensional position vector, and $\bm{k}=(k_x,k_y)$ denotes the two-dimensional transverse wavenumber vector. Then, $F(\bm{k})$ is modulated by the vortex phase plate, $H(\bm{k})=e^{\\i\theta}$ with an azimuthal angle $\theta$ at the Fourier plane and the image is obtained after the inverse Fourier transform by the second lens as follows:

\begin{equation}
    f_{\text{out}} (\bm{r})= \mathcal{F}^{-1}\{F(\bm{k})H(\bm{k})\}=f_\text{in}(\bm{r})*h(\bm{r}),
    \label{Eq2_51}
\end{equation}
where $*$ denotes the convolution and $h(\bm{r})=\frac{\ii}{2\pi}\frac{e^{\ii \theta}}{r^2}$ is the inverse Fourier transform of $H(\bm{k})$ and denotes the point spread function of the optical system with the vortex phase plate as the Fourier filter. The PAO, captured by the bright-field microscope, undergoes edge enhancement by the $4f$ system shown in Fig. \ref{4fsystem}, resulting in edge detection at the image plane. The detected edge intensity is given by $I=|f_{\text{out}} (\bm{r})|^2$. Further, simplifying Eq. (\ref{Eq2_51}), we get final equation as:
\begin{equation}
    I=|f_{\text{out}} (\bm{r})|^2= \left|-\ii\left(\frac{\partial}{\partial x}  + \ii\frac{\partial}{\partial y}\right) f_\text{in}(\bm{r})*\frac{1}{r}\right|^2,
    \label{Eq2_52}
\end{equation}
where $r = \sqrt{x^2 + y^2}$ and the partial derivative $\left(\frac{\partial}{\partial x} + \ii\frac{\partial}{\partial y}\right)$ will detect the edges of the object while convolution term $1/r$ doesn’t contribute to detect the edges, instead, it broadens the edges of the object. Because the vortex filter is scalar one, the edges are non-uniform in the image plane of the $4f$ system, as shown in Fig. \ref{4fsystem} \cite{zangpo2023edge}.

\begin{figure}[htp!]
 \centering
\includegraphics[width=\linewidth]{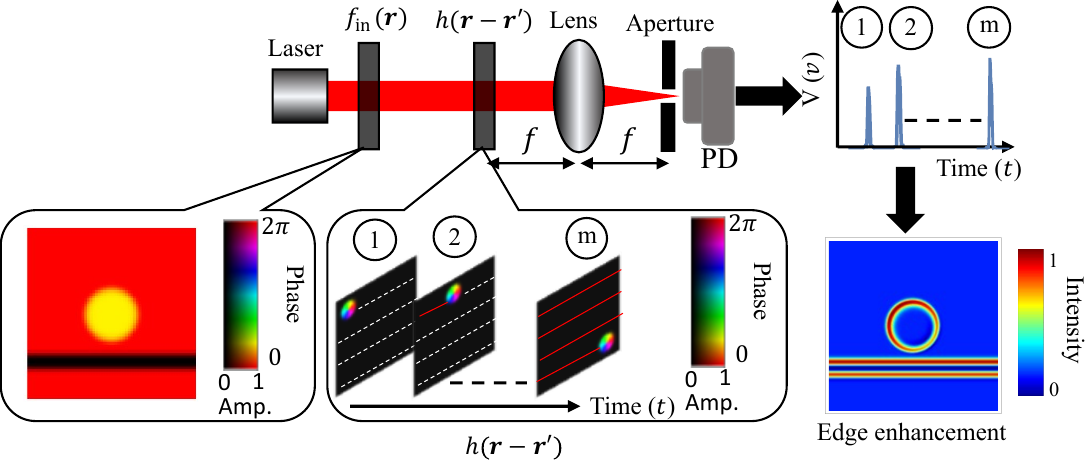}
\caption{Single-pixel edge-enhanced microscope via convolutional filtering. In $m$ different patterns, a white dot denotes the scanning path of patterns, while a red line denotes the patterns that have completed the scanning. For each scanning of a pattern, the photodiode measures the corresponding optical power as voltage data in time and the edge-enhanced image can be obtained by reshaping the temporal voltage data to a two-dimensional format.}
\label{fig5_1}
\end{figure}

In Eq. (\ref{Eq2_51}), $h(\bm{r})=\frac{\ii}{2\pi}\frac{e^{\ii \theta}}{r^2}$ consists of a spiral phase $e^{i \theta}$ and a localized amplitude function $a(\bm{r})\propto 1/r^2$. In general, the requirement of the filter function for the edge enhancement is the localized amplitude function multiplied by the azimuthal phase, i.e.,
$h(\bm{r})=a(\bm{r}) e^{i \theta}$. The convolution of $f_{\text{in}} (\bm{r})*h(\bm{r})$ can be rewritten as 
\begin{equation}
   f_{\text{out}} (\bm{r} ) = \iint_{-\infty}^{\infty} f_\text{in}(\bm{r'}) h(\bm{r}-\bm{r'})  d \bm{r'}=\mathcal{F}\{ f_\text{in}(\bm{r'}) h(\bm{r}-\bm{r'})\} (0),
    \label{Eq5_1}
\end{equation}
where the object $f_{\text{in}}(\bm{r})$ multiplies with shifted patterns of $h(\bm{r})$ and is then Fourier transformed, followed by extracting the center amplitude. In Eq. (\ref{Eq5_1}), (0) denotes the center pixel of the output. Equation (\ref{Eq5_1}) can be implemented as shown in Fig. \ref{fig5_1}, where the object light is multiplied with the shifted localized vortex phase $h(\bm{r}-\bm{r'})$ generated by the super-pixel method using the DMD, it is then Fourier transformed by the Fourier lens in the $2f$ system, followed by extracting the desired location with an aperture. This aperture has two functions: the first is to select first-order diffraction to generate the localized vortex phase by the super-pixel method, as shown later; the second function is to select the center position according to Eq. (\ref{Eq5_1}). Finally, the single-pixel detector, such as a photodiode, measures the corresponding optical power for different patterns over time to reconstruct the enhanced edges of the object by reshaping one-dimensional temporal data to a two-dimensional format, as shown in Fig. 2 with the caption ’Edge Enhancement’.

\subsection{Generation of localized vortex phase by super-pixel method with DMD}

Unlike a spatial light modulator (SLM), which can generate phase and amplitude holograms\cite{lightcontrol1,lightcontrol2,lightcontrol3,lightcontrol4,lightcontrol5}, the DMD can only generate a binary amplitude hologram represented by binary reflections with 1 or 0 indicating turn on and off, respectively. In the super-pixel method\cite{ampPhasewavefront}, however, both the amplitude and phase modulation of a light wave can be implemented by extracting the first-order diffraction generated by the binary amplitude hologram. This method offers higher beam shaping fidelity than SLMs\cite{DMDvsSLM1} and much faster refresh rates over 10 kHz for binary amplitude patterns, whereas most SLMs are limited to 120 Hz\cite{DMDvsSLM2,highspeed}. However, these advantages come at the expense of limited modulation depth and low diffraction efficiency\cite{DMDvsSLM1}.

\begin{figure}[htp!]
 \centering
\includegraphics[width=\linewidth]{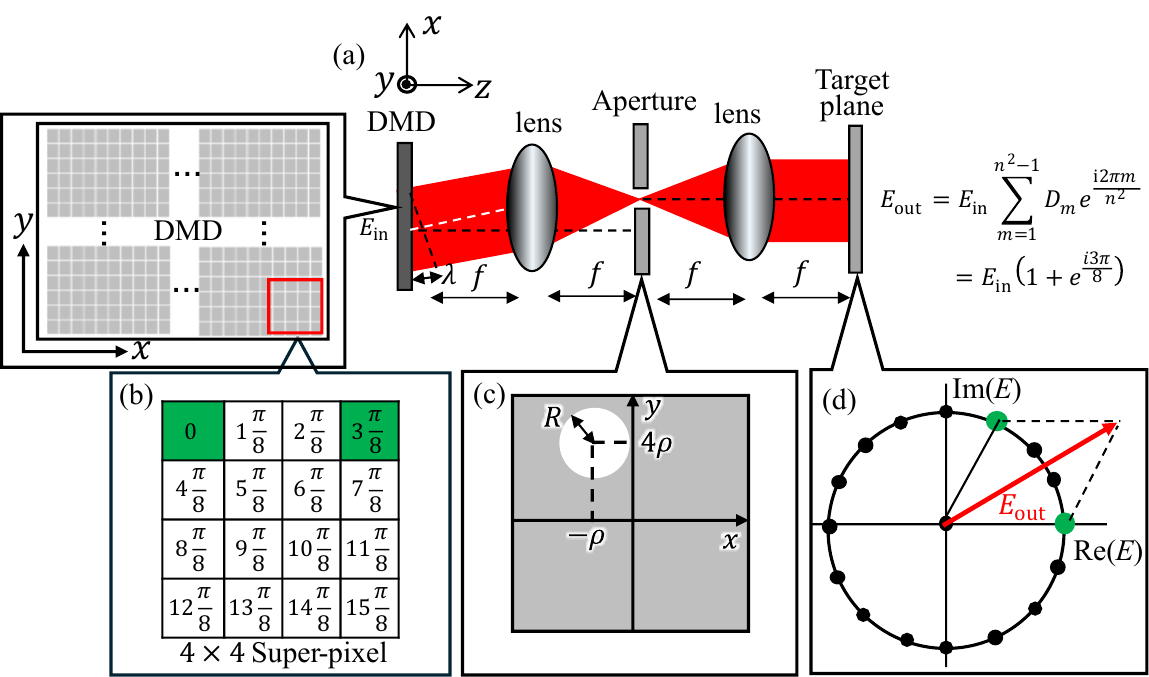}
\caption[Super-pixel method]{ (a) Setup of super-pixel method. (b) Super-pixel with a size of $4 \times 4$ pixels exhibiting a phase difference of $2\pi/n^2$ in the $x$-direction and $2\pi/n$ in the $y$-direction. Two pixels in green are turned on with 100\% reflectance while the other pixels are turned off with 0\% reflectance. (c) Aperture with radius of $R$ placed at $(-\rho, 4 \rho)$ on the Fourier plane to extract the first-order diffraction of DMD. (d) The resultant field $E_\text{out}$ (red arrow) on the target plane is proportional to the sum of the phases on the pixels that are turned on.}
\label{fig5_2}
\end{figure}

Figure \ref{fig5_2} (a) shows a schematic $4f$ setup of the super-pixel method to obtain full spatial control over the phase and amplitude of light on the target plane\cite{superpixel1}. The DMD placed at the front focal plane of the first lens is partitioned into super-pixels, which are square sub-groups of $n \times n$ micromirrors. The diffracted light from the DMD is Fourier transformed by the first lens and an aperture with a radius of $R$ acts as a spatial filter on the Fourier plane to capture the first-order diffraction. The position of the aperture is selected to ensure that the responses of adjacent pixels within the super-pixel on the DMD exhibit a phase difference of  $2\pi/n^2$ in the $x$-direction and $2\pi/n$ in the $y$-direction, as shown in Fig. 3 (b), where some pixels are highlighted in green to indicate that they are turned on. Figure \ref{fig5_2} (c) shows the appropriate aperture
position, which is obtained as $(x,y)=(-\rho, n \rho)$, where $\rho=\frac{-\lambda f}{n^2 d}$, $\lambda$ is the wavelength of the light, $f$ is the focal length of the first lens and $d$ is pixel size of the micromirror. The asymmetric phase variation between the $x$ and $y$ directions arises directly from this chosen aperture position\cite{superpixel1}. Here we notice that optimizing the aperture radius $R$ is crucial because a small $R$ will reduce the desired signal and a large $R$ will contain an undesired zero-order diffraction. The first-order diffraction extracted by the aperture further undergoes an inverse Fourier transform by the second lens. The resulting field $E_\text{out}$ at the target plane is proportional to the sum of valid phase contributions from individual pixel responses, which are evenly distributed on the complex plane along a circle, as shown in Fig. \ref{fig5_2} (d). To determine the resultant electric field $E_\text{out}$ at the target plane, we use the following equation:
\begin{equation}
  E_\text{out}=E_\text{in}\sum_{m=0}^{n^2-1} D_m e^{\ii 2\pi m/n^2}
   \label{esuperpixel},
\end{equation}
where $E_\text{in}$, $D_m$ and n denote the input electric field, the binary values 0 or 1 of the $m$-th pixel within the super-pixel, and the super-pixel size, respectively. As shown in Fig. 2, the second $2f$ system in Fig. \ref{fig5_2} (a) is not required in the proposed single-pixel edge enhancement, due to the Fourier transform in Eq. (\ref{Eq5_1}).
 
\begin{figure}[htp!]
 \centering
\includegraphics[width=0.9\linewidth]{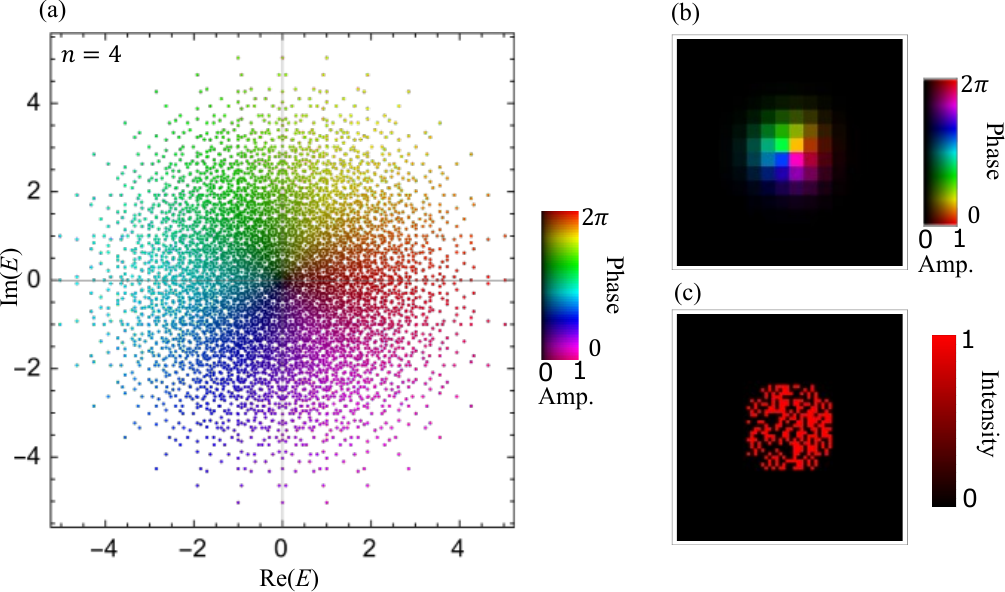}
\caption{(a) Complex amplitude generated by all possible combinations of pixels corresponding to the super-pixel with $n=4$. (b) The desired complex amplitude distribution with $16 \times 16$ pixels. (c) The hologram with $64 \times 64$ pixels generated by replacing super-pixel $4 \times 4$ to the $16 \times 16$ desired complex function.}
\label{complextarget}
\end{figure}

In Fig. \ref{complextarget} (a), we plot the complex amplitudes for all possible combinations of $D_m$ within a super-pixel of a size $n= 4$. As can be seen from Fig. \ref{complextarget} (a), the possible complex amplitude is distributed within a disk with a radius of $\text{cot}(\pi/n^2)\approx5$ on the complex plane. In the set of 65536 patterns, there are many duplicated complex fields and only 6561 patterns have unique fields. For the purpose of making a hologram using the super-pixel method, each pixel in the desired complex function is replaced with an appropriate super-pixel of a size $n \times n$. The replacement process involves comparing the resultant field of the super-pixel with each pixel in the desired complex function and the nearest values from all the possible fields generated by the super-pixel are then assigned to each pixel in the desired complex function. Figure \ref{complextarget} (b) illustrates a grid of $16 \times 16$ pixels of the desired function with the localized vortex phase. By employing the super-pixel method, each $4 \times 4$ super-pixel replaces individual pixels in the desired complex function in Fig. \ref{complextarget} (b), resulting in a hologram pattern as shown in Fig. \ref{complextarget} (c) with a dimension of $64 \times 64$ pixels.

\section{Numerical Simulation Results of the Proposed Method}
\label{sec:5_3}  
Under numerical simulation, there are three sections. First, we investigate the edge enhancement capability of localized amplitude functions. Secondly, we show the fidelity estimation of the super-pixel method to generate the localized vortex phase. In the final section, we demonstrate the edge enhancement capability of single-pixel imaging via convolutional filtering with the localized vortex phase.

\subsection{Edge enhancement capability of localized amplitude functions}
\label{traditional}
\begin{figure}[htp!]
 \centering
\includegraphics[width=\linewidth]{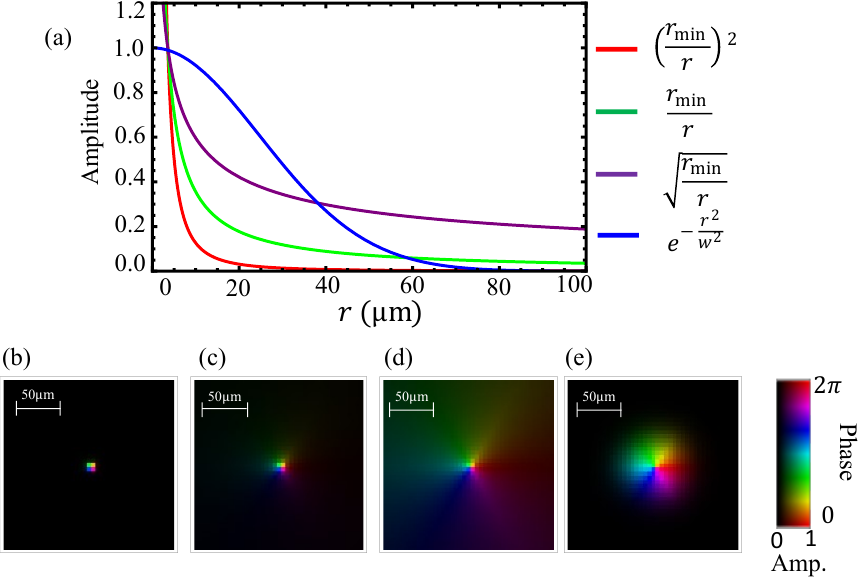}
\caption{(a) Radial dependence of the amplitude functions, $\frac{1}{r^2}$, $\frac{1}{r}$, $\frac{1}{\sqrt{r}}$ and $e^{-r^2/w^2}$ with $w=35$ \textmu m. (b-e) Corresponding complex amplitude distributions of the localized vortex phase.}
\label{fig5_3}
\end{figure}

In this section, we consider several types of localized amplitude functions $a(\bm{r})$, such as $\frac{1}{r^2}$, $\frac{1}{r}$, $\frac{1}{\sqrt{r}}$ and $e^{-r^2/w^2}$ with the radius $w$, as plotted in Fig. \ref{fig5_3} (a), where we use the optimized radius $w=35$ \textmu m for the Gaussian function, explained in Section \ref{simulationresult}. The complex amplitude distribution of $h(\bm{r})=a(\bm{r})e^{\ii \theta}$ with several localized amplitude functions $a(\bm{r})$ are shown in Figs. \ref{fig5_3} (b) - (e). Here we note that $\frac{1}{r^2}$, $\frac{1}{r}$, and $\frac{1}{\sqrt{r}}$ in Fig. \ref{fig5_3} are normalized by their function values at $r_\text{min}$, which is the minimum radial value in the simulation coordinate.

\begin{figure}[htp!]
 \centering
\includegraphics[width=.80\linewidth]{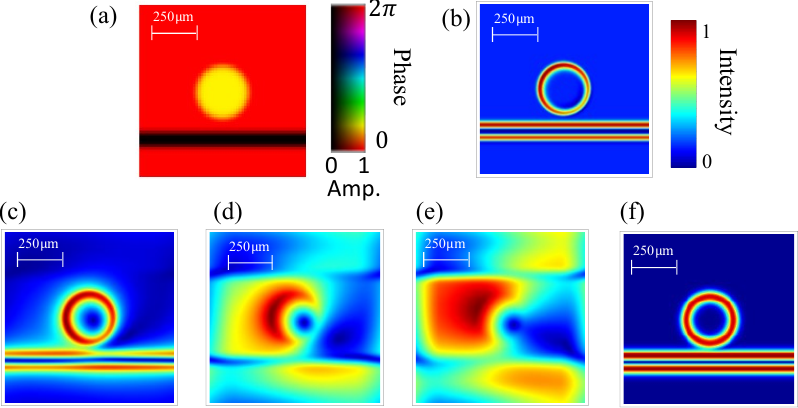}
\caption{Edge enhancement capability of several localized amplitude functions. (a) Complex amplitude distribution of the PAO. (b) Ideal edge enhancement. (c-f) Edge enhancement using $\frac{1}{r^2}$, $\frac{1}{r}$, $\frac{1}{\sqrt{r}}$, and $e^{-r^2/w^2}$ with $w=35$ \textmu m in $h(\bm{r})$, respectively.}
\label{traditionaledge}
\end{figure}

Before using the super-pixel method, we use Eq. ({\ref{Eq2_51}}) to numerically verify the edge enhancement capability of the localized amplitude functions. Here we consider the PAO in Fig. \ref{traditionaledge} (a) as the object. We first perform an inverse Fourier transform on $h(\bm{r})$ to obtain $H(\bm{k})$ and then perform a Fourier transform on the PAO, followed by multiplication with $H(\bm{k})$. Finally, we apply an inverse Fourier transform to the Fourier spectrum and then take the absolute square to obtain the edges of the PAO. The simulation parameters are as follows: a wavelength of 635 nm, a total dimension of $1\text{mm} \times 1\text{mm}$, a step size of $1/199 \text{mm} \times 1/199 \text{mm}$, which corresponds to  $200 \times 200$ pixels.

Figure \ref{traditionaledge} shows the numerical simulation results of the edge enhancement using the localized vortex phase $h(\bm{r})$. Figure \ref{traditionaledge} (b) shows the ideal edge enhancement of PAO by using $H(\bm{k})=k_r e^{\ii \theta}$ with ${k}_{{r}}\equiv\sqrt{k_x^2+k_y^2}$, while Figs. \ref{traditionaledge} (c-f) illustrate edge enhancement using the localized amplitude functions $a(\bm{r})$. The localized amplitude functions $\frac{1}{r}$ and $\frac{1}{\sqrt{r}}$ fail to detect the edges of the PAO while the functions $\frac{1}{r^2}$ and $e^{-r^2/w^2}$ successfully detect the edges of the PAO. The correlation coefficients between ideal edge enhancement in Fig. \ref{traditionaledge} (b) and Figs. \ref{traditionaledge} (c-f) are 0.80, 0.25, 0.06 and 0.85, respectively. The reason why the amplitude functions $\frac{1}{r}$ and $\frac{1}{\sqrt{r}}$ of optical vortex fail to detect the edges of object is that their radial dependence is not tightly localized compared to $\frac{1}{r^2}$ and $e^{-r^2/w^2}$, i.e., the tails of the radial dependence for the amplitude functions $\frac{1}{r}$ and $\frac{1}{\sqrt{r}}$ do not rapidly approach zero, instead, they exhibit a higher value, as can be seen in Fig. \ref{fig5_3} (a).

\subsection{Fidelity estimation of super-pixel method to generate localized vortex phase}
\label{fidelitysimu}
The modulation accuracy of the super-pixel method between target fields and desired fields can be quantified by calculating the fidelity, $F$, as shown below\cite{superpixel1}:

\begin{equation}
  F = \frac{|\bm{D}_{\text{desired}}^{*} \cdot \bm{D}_{\text{out}}|^2}{|\bm{D}_{\text{desired}}|^2 |\bm{D}_{\text{out}}|^2},
   \label{fidelityl}
\end{equation}
where $*$ denotes the complex conjugate. The $\bm{D}_{\text{out}}$ and $\bm{D}_{\text{desired}}$ are one column vectors reshaped from the two-dimensional target fields, $E_{\text{out}}(\bm{r})$, and desired fields, $E_{\text{desired}}(\bm{r})$, respectively. When $F=1$, it indicates that the similarity between target fields and desired fields is the same, whereas when $F=0$, it means they are completely different.

We conduct a numerical simulation to estimate the fidelity between the target field generated by the super-pixel method and the desired localized vortex phase with the amplitude functions, $(c/r)^2$, $c/r$, $\sqrt{c/r}$ and $c e^{-r^2/w^2}$, where $c$ denotes the amplitude coefficient. The optimizing parameters to obtain the best fidelity in the super-pixel method are the amplitude coefficient $c$ and the aperture radius $R$. The optimal values were determined by a systematic parameter sweep: the fidelity $F$ was calculated for a wide range of each parameter $R/\rho$ from 0 to 4 and for the Gaussian function, the waist $w$ was also varied from 0 to 100 \textmu m. The combination of values yielding the maximum fidelity $F$ was selected. We use $w=100$ \textmu m, which gives the highest fidelity of $0.89$. The simulation conditions remain the same as those in Section \ref{traditional}, with an additional parameter, a focal length of the lens setting at 300 mm.

\begin{figure}[htp!]
 \centering
\includegraphics[width=\linewidth]{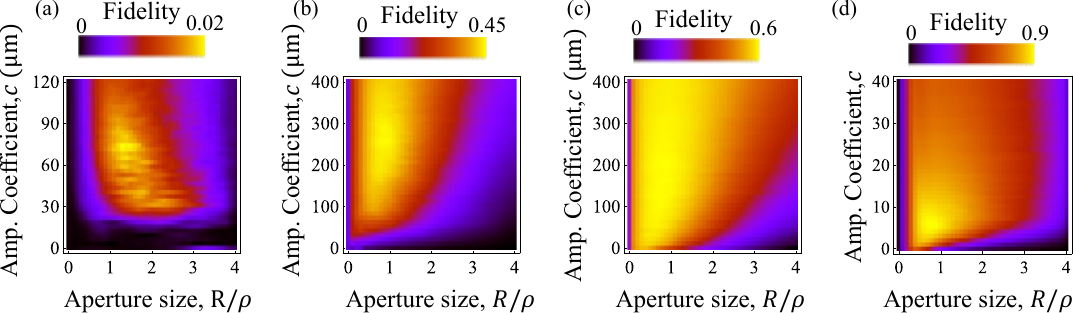}
\caption{(a-d) The fidelity of the localized vortex phase with amplitude functions:$\frac{1}{r^2}$, $\frac{1}{r}$, $\frac{1}{\sqrt{r}}$ and $e^{-r^2/w^2}$ ($w=100$ \textmu m) for different amplitude coefficients $c$ and aperture radius, $R$ normalized by aperture position $\rho$.}
\label{fig5_4}
\end{figure}

Figures \ref{fig5_4} (a-d) show the simulation results of the fidelity between the ideal and the generated localized vortex phases with different $a(\bm{r})$: $(c/r)^2$, $c/r$, $\sqrt{c/r}$ and $c e^{-r^2/w^2}$, respectively. The horizontal and vertical axes in Fig. \ref{fig5_4} show the aperture radius $R$ normalized by aperture position $\rho$ and the amplitude coefficient $c$ of $h(\bm{r})$, respectively. Table \ref{tab:fidelity} shows a summary of the maximum fidelities and the corresponding optimized parameters. The Gaussian amplitude function $e^{-r^2/w^2}$ provides a fidelity of $0.89$, whereas $\frac{1}{r^2}$, $\frac{1}{r}$ and $\frac{1}{\sqrt{r}}$ yield fidelity values below $0.6$ (refer to Table \ref{tab:fidelity}), because these power functions increase rapidly as they approach the origin, resulting in low fidelity when implemented with the finite amplitude range of the super-pixel method. Even if the fidelity to the desired function is low, however, edge enhancement might be possible if a certain degree of localization is achieved by the amplitude function. Thus, in the next section, we use the amplitude functions $\frac{1}{r^2}$ and $e^{-r^2/w^2}$ which successfully perform edge enhancement in Fig. \ref{traditionaledge}, to evaluate our single-pixel method via convolutional filtering.

\begin{table}[!ht]
\centering
\caption{Simulation results of maximum fidelity and optimized parameters with
different $a(\bm{r})$}
\label{tab:fidelity}
\begin{tabular}{cccccc}
 \hline
\backslashbox{Parameter}{$a(\bm{r})$} & $\frac{1}{r^2}$ & $\frac{1}{r}$ & $\frac{1}{\sqrt{r}}$ & $e^{-r^2/w^2}$ \\ 
 \hline
Max. fidelity, $F$ & 0.02 & 0.45 & 0.57 & 0.89\\ 
 \hline
Radius, $R/\rho$ & 2.5 & 2.6 & 2 & 0.7\\ 
 \hline
Amp. coefficient, $c$  & 42 \textmu m & 100 \textmu m & 100 \textmu m & 9 \\ 
 \hline
\end{tabular}
\end{table}

\subsection{Edge enhancement capability of single-pixel imaging via convolutional filtering
with localized vortex phase}
\label{simulationresult}
Now we evaluate the edge enhancement capability of our proposed single-pixel imaging method
with the localized amplitude functions, $\frac{1}{r^2}$ and $e^{-r^2/w^2}$. We first prepared the binary hologram on the DMD to generate the shifted localized vortex phase $h(\bm{r}'-\bm{r})$ by using the super-pixel method with $n=4$. By following the principles of Section \ref{proposed method}, we calculated the Fourier transform of the object light multiplied with the hologram on the DMD and then obtained optical power at the center point of the aperture, $(x,y)=(-\rho, n \rho)$, to reconstruct the two-dimensional edge enhanced image. The simulation conditions remained the same as those in Section \ref{traditional}.

\begin{figure}[htp!]
 \centering
\includegraphics[width=0.8\linewidth]{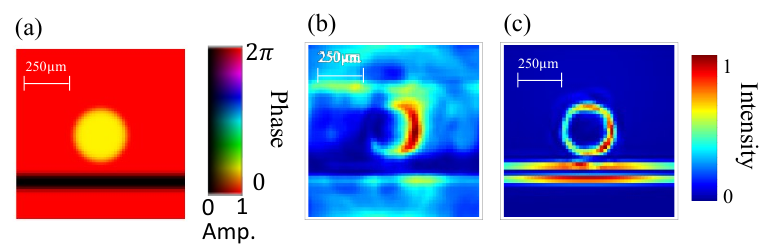}
\caption[Numerical simulation: Edge enhancement using the proposed method.]{(a) Distribution of PAO, (b) and (c) Edge enhancement via proposed method using the optical vortex with the amplitude function $1/r^2$ and $e^{-r^2/w^2}$ with $w=35$ \textmu m, respectively.}
\label{fig5_5}
\end{figure}

\begin{figure}[htp!]
 \centering
\includegraphics[width=0.4\linewidth]{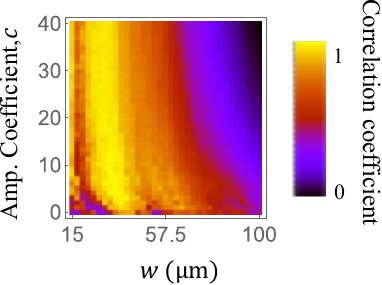}
\caption{Correlation coefficient between the edge enhanced by the Gaussian amplitude function of the optical vortex and the ideal edge enhancement for varying values of $w$ and amplitude coefficient $c$.}
\label{optimizeEdgeCC}
\end{figure}

Figure \ref{fig5_5} (a) represents the complex amplitude distribution of the PAO. The ideal edge to compare with the edge enhancement of the object by vortex phases with the amplitude function $1/r^2$ and $e^{-r^2/w^2}$ is the same as Fig. \ref{traditionaledge} (b) but with a different pixel size i.e., $50 \times 50$, which represents the total pixel size for the single-pixel edge enhancement reconstruction. The best edge enhancement using localized vortex phases with the amplitude function $1/r^2$ and $e^{-r^2/w^2}$ are shown in Figs. \ref{fig5_5} (b) and (c), respectively. The amplitude function $e^{-r^2/w^2}$ could successfully enhance the edge of the object using the proposed single-pixel edge enhancement technique, while the amplitude function $1/r^2$, which could successfully enhance the edge in Section \ref{traditional}, failed to enhance the edge. The main reason of this failure is considered to be low fidelity (0.02), as shown in Table \ref{tab:fidelity}. Figure \ref{optimizeEdgeCC} shows the dependence of the correlation coefficient between the single-pixel edge enhancement by using the Gaussian amplitude function and the ideal edge enhancement on the optimizing parameters $w$ and $c$. We selected a starting point of $w$ greater than 15 \textmu m, which exceeded the radius of the super-pixel size of 10 \textmu m, because $w$ values lower than 15 \textmu m did not detect the edges of the object. The highest correlation coefficient 0.95 was achieved when $w=35$ \textmu m and $c=6$. The edge enhancement using these optimized parameters is illustrated in Fig. \ref{fig5_5} (c). The strong and positive relationship between the ideal edge and the edge produced by the proposed method demonstrates its effectiveness in enhancing the edges of objects.

\section{Conclusion}
\label{sec:5_6}
In this paper, we proposed and numerically demonstrated the edge enhancement of objects using single-pixel imaging via convolutional filtering with a localized vortex phase (spiral phase $e^{i\theta}$ and amplitude function $a(\bm{r})$). Through systematic testing of localized functions, the Gaussian amplitude achieved the highest fidelity (0.89) using the super-pixel method. Our simulations yielded a 0.95 correlation coefficient between ideal and enhanced edges, confirming the method's ability to enhance phase-amplitude object (PAO) edges while eliminating the measurement overhead of phase-shifting approaches. This method shows strong potential for non-visible wavelength applications such as near-infrared fluorescence microscopy and silicon semiconductor circuit inspection, where conventional sensors face limitations.

Future work will focus on three key directions to further advance this methodology: (1) validation across diverse samples including biological specimens and semiconductor inspection patterns
to establish robustness in real-world applications,  (2) optimization of vortex generation through comparative analysis of the super-pixel method with alternative approaches such as error diffusion \cite{referee_1_errorDiffussion} and Direct Binary Search \cite{referee_1_DBS} to evaluate speed-fidelity trade-offs, and (3) single-pixel edge enhancement with randomly-distributed localized vortex filter for faster acquisition time.

\backmatter

\section*{Declarations}
\begin{itemize}
\item Funding

Japan Society for the Promotion of Science (18KK0079, 20K05364); Research Foundation for Opto-Science and Technology.
\item Conflict of interest/Competing interests 

The authors declare that they have no known competing financial interests or personal relationships that could have appeared to influence the work reported in this paper.
\item Ethics approval and consent to participate

Not applicable.
\item Consent for publication

Not applicable.
\item Data availability 

Data underlying the results presented in this paper are not publicly available at this time but may be obtained from the authors upon reasonable request.

\item Author contribution

All authors contributed equally.
\end{itemize}

\bibliography{sn-bibliography}

\end{document}